\documentclass[sigconf,nonacm]{acmart}
\AtBeginDocument{%
  }
\copyrightyear{2025}
\acmYear{2025}
\setcopyright{rightsretained}
\acmConference[IMC '25]{Proceedings of the 2025 ACM Internet Measurement Conference}{October 28--31, 2025}{Madison, WI, USA}
\acmBooktitle{Proceedings of the 2025 ACM Internet Measurement Conference (IMC '25), October 28--31, 2025, Madison, WI, USA}
\acmDOI{10.1145/3730567.3768601}
\acmISBN{979-8-4007-1860-1/2025/10}

\usepackage{enumitem}
\usepackage{subcaption}

\begin{document}

\title{Poster: ChatIYP: Enabling Natural Language Access to the Internet Yellow Pages Database}


% \author{Vasilis Andritsoudis}
% \affiliation{
%   \institution{Aristotle University of Thessaloniki}
%   \city{Thessaloniki}
%   \country{Greece}
% }

% \author{Pavlos Sermpezis}
% \affiliation{
%   \institution{Aristotle University of Thessaloniki}
%   \city{Thessaloniki}
%   \country{Greece}
% }

% \author{Ilias Dimitriadis}
% \affiliation{
%   \institution{Aristotle University of Thessaloniki}
%   \city{Thessaloniki}
%   \country{Greece}
% }

% \author{Athina Vakali}
% \affiliation{
%   \institution{Aristotle University of Thessaloniki}
%   \city{Thessaloniki}
%   \country{Greece}
% }
\author{Vasilis Andritsoudis}
\affiliation{%
  \institution{Aristotle University of Thessaloniki}
  \city{Thessaloniki}
  \country{Greece}
}
\email{vasandven@csd.auth.gr}

\author{Pavlos Sermpezis}
\additionalaffiliation{%
  \institution{Aristotle University of Thessaloniki}
  \city{Thessaloniki}
  \country{Greece}
}
\affiliation{%
  \institution{Measurement Lab\\Code for Science \& Society }
  \city{Portland}
  \state{OR}
  \country{ United States}
}
\email{pavlos@measurementlab.net}

\author{Ilias Dimitriadis}
\affiliation{%
  \institution{Aristotle University of Thessaloniki}
  \city{Thessaloniki}
  \country{Greece}
}
\email{idimitriad@csd.auth.gr}

\author{Athena Vakali}
\affiliation{%
  \institution{Aristotle University of Thessaloniki}
  \city{Thessaloniki}
  \country{Greece}
}
\email{avakali@csd.auth.gr}

\renewcommand{\shortauthors}{Vasilis Andritsoudis et al.}

\begin{abstract}
The Internet Yellow Pages (IYP) aggregates information from multiple sources about Internet routing into a unified, graph-based knowledge base. However, querying it requires knowledge of the Cypher language and the exact IYP schema, thus limiting usability for non-experts. In this paper, we propose \textit{ChatIYP}, a domain-specific Retrieval-Augmented Generation (RAG) system that enables users to query IYP through natural language questions. Our evaluation demonstrates solid performance on simple queries, as well as directions for improvement, and provides insights for selecting evaluation metrics that are better fit for IYP querying AI agents.
\end{abstract}

%    <concept>
%        <concept_id>10003033.10003079.10011704</concept_id>
%        <concept_desc>Networks~Network measurement</concept_desc>
%        <concept_significance>500</concept_significance>
%        </concept>
%    <concept>
%        <concept_id>10003120.10011738.10011776</concept_id>
%        <concept_desc>Human-centered computing~Accessibility systems and tools</concept_desc>
%        <concept_significance>500</concept_significance>
%        </concept>
%  </ccs2012>

% \ccsdesc[500]{Networks~Network measurement}
% \ccsdesc[500]{Human-centered computing~Accessibility systems and tools}

% \keywords{IYP, RAG, Internet, LLM}

% \received{14 June 2025}
% \received[revised]{14 June 2025}
% \received[accepted]{14 June 2025}

\maketitle

\section{Introduction} \label{sec:introduction}

Information about the Internet's underlying infrastructure and routing (e.g., ASes, IP prefixes) is valuable for researchers and engineers, for, e.g., diagnosing routing anomalies. However, this data is often fragmented across disparate sources (e.g., BGP routing tables, WHOIS records). The \textit{Internet Yellow Pages (IYP)}~\cite{fontugne2024wisdom} consolidates this data into a unified, graph-based knowledge base. It models infrastructure entities (e.g., ASes, IP blocks) as nodes, with edges capturing relationships. This rich structure supports powerful queries, but only through Cypher~\cite{cypher}, a graph query language that requires knowledge of graph syntax and schemas. For example, answering a simple question like:

\begin{center}
\textit{``What is the percentage of Japan's population in AS2497?''}    
\end{center}

\noindent requires a Cypher query such as:

\begin{center}
\texttt{MATCH (:AS \{asn:2497\})-[p:POPULATION]-(:Country \{country\_code:'JP'\}) RETURN p.percent}
\end{center}

\noindent This complexity creates a steep learning curve, limiting IYP’s adoption beyond expert users. 

To make IYP easily for everyone, in this paper, we design and implement \textit{ChatIYP}, a system build on a domain-specific Retrieval-Augmented Generation (RAG)~\cite{lewis2020retrieval} approach, which translates natural language questions into Cypher queries, executes them on IYP, and returns both the lexical responses and the underlying query for transparency.

We evaluate ChatIYP using multiple example queries and evaluation metrics. Results show that responses are accurate in most easy technical tasks; while there is room for improvement for harder tasks. Moreover, we demonstrate that an evaluation framework using LLM-as-a-judge setup (G-Eval~\cite{liu2023g}) better reflects human judgment in query quality compared to other common metrics for RAG evaluation.

% By lowering the barrier to querying Internet infrastructure data, ChatIYP enables more intuitive and widespread use of graph-based insights in networking research and operations.

\section{ChatIYP} \label{sec:chatiyp}

\begin{figure}[t]
\centering
\includegraphics[width=0.45\linewidth]{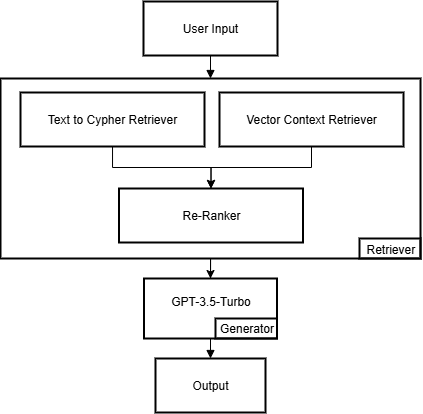}
\caption{ChatIYP architecture pipeline.}
\label{fig:chatiyp}
\end{figure}

% We develop \textbf{ChatIYP}, a retrieval-augmented LLM-based assistant for answering user questions over the IYP graph. The system outputs both a natural language answer and the corresponding Cypher query, supporting transparency and enabling downstream structured processing.

% The key challenge is the gap between user-friendly queries and the structured, large-scale IYP graph. Encoding the graph directly into an LLM is infeasible. Instead, ChatIYP uses a lightweight \textit{RAG} pipeline to query the graph on demand and guide generation with retrieved context.

Figure~\ref{fig:chatiyp} depicts the RAG-based architecture of ChatIYP, which consists of three key stages:

\noindent \textbf{(1) User Query}: A user submits a natural language question through a web interface; this is the input for our system.

\noindent \textbf{(2) Retrieval}: ChatIYP retrieves relevant information from the IYP graph database using three complementary ways, combining symbolic and semantic methods
\begin{itemize}[left=0pt]
    \item \textit{TextToCypherRetriever:} An LLM maps the user query to a Cypher query. We designed a prompt chain fine-tuned on IYP query patterns. The resulting query is executed against the Neo4j graph to return structured subgraphs.
    \item \textit{VectorContextRetriever:} When structured queries fail or sparse results are returned, dense embeddings for node descriptions are used to fetch textual context of nearby graph nodes (via vector similarity). This is particularly useful for vague queries or cases where graph structure alone is insufficient.  
    \item \textit{LLMReranker:} Given multiple retrieval candidates from the above steps, we re-rank results using a shallow LLM-based scorer to improve context selection before generation. This combination provides robustness: when symbolic translation fails or yields low recall, semantic retrieval ensures we still return useful information.
\end{itemize}

\noindent \textbf{(3) Generation}: The input and the retrieved nodes are passed to an LLM that generates a natural language response and a refined Cypher query. This is the output of the system that is displayed to the user. In our implementation and experimenation, we use GPT-3.5-Turbo as the backbone LLM for response generation

We implement the pipeline using the LlamaIndex~\cite{llamaindex} framework, which supports symbolic and semantic retrieval over graph-structured data and provides built-in integration with Neo4j~\cite{neo4j}. %Unlike more complex frameworks such as LangChain, LlamaIndex offers fast iteration cycles and a minimal setup footprint, which aligns with our focus on rapid prototyping and scalability testing.

\section{Results} \label{sec:results}

We evaluate ChatIYP using the \textit{CypherEval}~\cite{cyphereval} dataset, a benchmark of more than 300 natural language questions over IYP. Each question is annotated with a gold Cypher query and labeled by difficulty: \textit{Easy}, \textit{Medium}, or \textit{Hard}, across both \textit{general} and \textit{technical} domains.

\begin{figure}[t]
\centering
\begin{subfigure}[t]{0.48\linewidth}
    \centering
    \includegraphics[width=\linewidth]{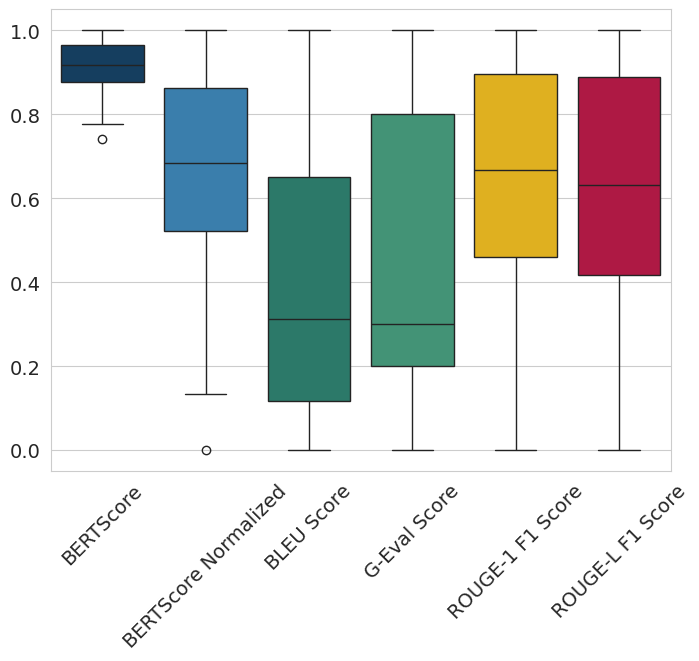}
    % \caption{Comparison of metric distributions.}
    \caption{}
    \label{fig:metrics_boxplot}
\end{subfigure}
\hfill
\begin{subfigure}[t]{0.5\linewidth}
    \centering
    \includegraphics[width=\linewidth]{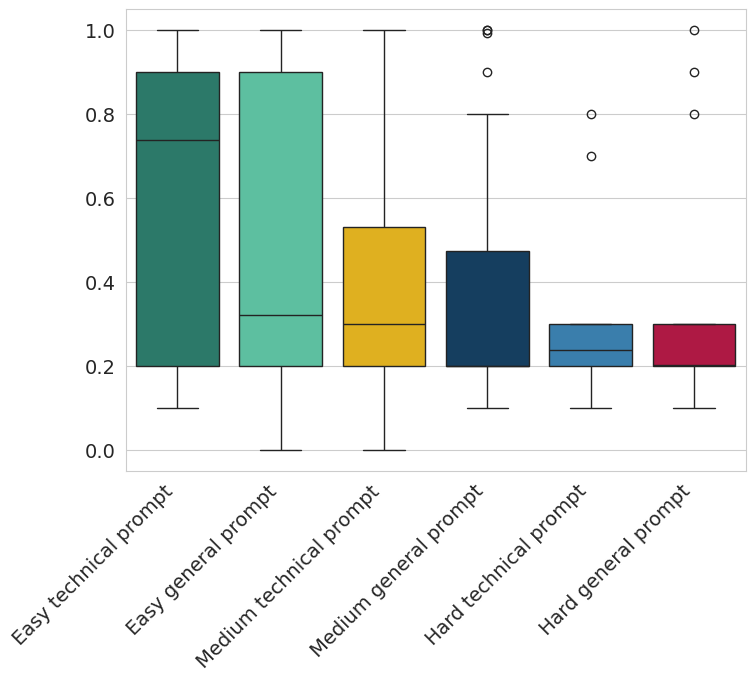}
    % \caption{G-Eval scores by difficulty.}
    \caption{}
    \label{fig:geval_difficulty_boxplot}
\end{subfigure}
\caption{(a) Comparison of metric distributions. (b) G-Eval scores by difficulty.}
\label{fig:combined_boxplots}
\end{figure}

\noindent\textbf{Evaluation Setup.} To assess response quality, we use a validation model that executes the gold Cypher query on the IYP graph and prompts GPT-3.5 to produce a reference answer. ChatIYP’s output, is compared against these references using widespread text generation metrics. Figure~\ref{fig:metrics_boxplot} shows the results of this evaluation.

\noindent\textbf{Evaluation Metrics}. We apply several widely used metrics in natural language generation to quantify answer quality:
\begin{itemize}[left=0pt]
    \item \textit{BLEU}\cite{papineni2002bleu} evaluates n-gram precision between the model’s response and the reference.
    \item \textit{ROUGE}\cite{lin2004rouge} focuses on n-gram and subsequence recall.
    \item \textit{BERTScore}\cite{zhang2019bertscore} measures similarity using contextual embeddings from a pre-trained language model (e.g., BERT).
    \item \textit{G-Eval}\cite{liu2023g} uses GPT-4 as a judge to assess responses on factuality, relevance, and informativeness.
\end{itemize}

\noindent\textbf{Finding 1: G-Eval outperforms traditional metrics.} Standard metrics like BLEU, ROUGE, or even BERTScore underperform in this setting, primarily due to their reliance on surface-level token overlap. These metrics struggle with rephrased or factually incorrect answers. We observe that: (i) BLEU scores are overly penalized by minor phrasing mismatches, despite semantic correctness. (ii) ROUGE better accommodate reworded answers, but still correlate poorly with factual accuracy. (iii) BERTScore exhibits a ceiling effect that blurs performance distinctions, especially in responses with narrow linguistic variation (as common in IYP queries).

On the contrary, G-Eval has a bimodal score distribution, which provides clear separation between good and bad responses, aligning closely with human judgment. We therefore adopt G-Eval as our primary evaluation metric.

\noindent\textbf{Finding 2: ChatIYP handles simple prompts well}. Figure~\ref{fig:geval_difficulty_boxplot} presents G-Eval scores across difficulty categories. ChatIYP performs well on easy prompts, with over half of responses scoring above 75\%. Performance degrades with prompt complexity, particularly on hard questions involving multi-hop reasoning. Interestingly, no consistent performance gap emerges between general and technical prompts, suggesting that \textit{structural complexity, not domain specificity, poses the greatest challenge}.

% \begin{figure}[t]
% \centering
% \includegraphics[width=0.5\linewidth]{Figures/metrics_boxplot.png}
% \caption{Comparison of metric distributions.}
% \label{fig:metrics_boxplot}
% \end{figure}

% \begin{figure}[t]
% \centering
% \includegraphics[width=0.5\linewidth]{Figures/geval_difficulty_boxplot.png}
% \caption{G-Eval scores by difficulty.}
% \label{fig:geval_difficulty_boxplot}
% \end{figure}

\section{Conclusion} \label{sec:conclusion}

ChatIYP introduces an approach for a retrieval-augmented natural language interface for the IYP graph, thus simplifying access to complex network data. Initial evaulation shows strong performance on simple queries and underscores challenges with complex ones which opens the door for further future research. Additionally, the G-Eval metric better reflects human judgments than traditional metrics like BLEU or ROUGE. We provide a publicly accessible web application for ChatIYP at~\cite{chatiyp_webapp} and the full source code (including all evaluation results used in this paper) at~\cite{chatiyp_github}.

\bibliographystyle{ACM-Reference-Format}
\bibliography{Bibliography}

\end{document}